\documentclass[12pt]{article}

\usepackage[dvips]{graphicx}
\renewcommand{\baselinestretch}{1.75}
\font\scaps=cmcsc10    
\font\bbf=cmbx10 scaled\magstep4
\font\rmbig=cmr10 scaled\magstep2
\font\bigtensl=cmsl10 scaled\magstep2
\newcommand \be{\begin{equation}}
\newcommand \ee{\end{equation}}
\newcommand \ba{\begin{eqnarray}}
\newcommand \ea{\end{eqnarray}}
\begin{document}

\def\today{\ifcase\month\or
 January\or February\or March\or April\or May\or June\or
 July\or August\or September\or October\or November\or December\fi
 \space\number\day, \number\year}
%
\hfil PostScript file created: \today{}; \ time \the\time \ minutes

\centerline {EARTHQUAKE NUMBER FORECASTS TESTING
}

\vskip .15in
\begin{center}
{Yan Y. Kagan }
\end{center}
\centerline {Department of Earth, Planetary, and Space
Sciences, University of California,}
\centerline {Los Angeles, California 90095-1567, USA;}
\centerline {Emails: {\tt kagan@moho.ess.ucla.edu}}
\vskip 0.02 truein

\vspace{0.15in}


\noindent
{\bf Abstract.}
We study the distributions of earthquake numbers in two global
earthquake catalogs: Global Centroid-Moment Tensor (GCMT) and
Preliminary Determinations of Epicenters (PDE).
The properties of these distributions are especially required
to develop the number test for our forecasts of future
seismic activity rate, tested by the Collaboratory for
Study of Earthquake Predictability (CSEP).
A common assumption, as used in the CSEP tests, is that the
numbers are described by the Poisson distribution.
It is clear, however, that the Poisson assumption for the
earthquake number distribution is incorrect, especially for
the catalogs with a lower magnitude threshold.
In contrast to the one-parameter Poisson distribution so
widely used to describe earthquake occurrences, the
negative-binomial distribution (NBD) has two parameters.
The second parameter can be used to characterize the
clustering or over-dispersion of a process.
We investigate the dependence of parameters for both
distributions on the catalog magnitude threshold and on
temporal subdivision of catalog duration.
Firstly, we study whether the Poisson law can be statistically
rejected for various catalog subdivisions.
We find that for most cases of interest the Poisson
distribution can be shown to be rejected statistically at a
high significance level in favor of the NBD.
Therefore we investigate whether these distributions fit
the observed distributions of seismicity.
For this purpose we study upper statistical moments of
earthquake numbers (skewness and kurtosis) and compare them to
the theoretical values for both distributions.

Empirical values for the skewness and the kurtosis increase
for the smaller magnitude threshold and increase with even
greater intensity for small temporal subdivision of catalogs. 
As is known, the Poisson distribution for large rate values
approaches the Gaussian law, therefore its skewness and
kurtosis both tend to zero for large earthquake rates: for the
Gaussian law these values are identically zero.
A calculation of the NBD skewness and kurtosis levels based on
the values of the first two statistical moments of the
distribution, shows rapid increase of these upper moments
levels. 
However, the observed catalog values of skewness and kurtosis
are rising even faster. 
This means that for small time intervals the earthquake number
distribution is even more heavy-tailed than the NBD predicts.
Therefore for small time intervals we propose using empirical
number distributions appropriately smoothed for testing
forecasted earthquake numbers.

\vskip .15in
\noindent
{\bf Short running title}:
{\sc
Earthquake number forecasts
}

\vskip 0.05in
\noindent
{\bf Key words}:
\vskip .05in
Probability distributions;
Seismicity and tectonics;
Statistical seismology;
Dynamics: seismotectonics;
Subduction zones;

\newpage

\section{Introduction}
\label{intro}

This work is continuation of our study of earthquake number
distribution in earthquake catalogs (Kagan 2010; Kagan 2014;
Kagan and Jackson, 2016).
This time we conduct the investigations in a more rigorous
manner in order to create numerical guidelines for testing
forecasts of seismic activity similar to Bird et al. (2015).

We first briefly review discrete theoretical distributions
that are used to approximate the earthquake number
distribution (see Section~\ref{theo}).
These distributions are the Poisson and the negative-binomial
distribution (NBD).

To investigate the empirical pattern of earthquake occurrence
we study the distributions of earthquake numbers in two global
earthquake catalogs: Global Centroid-Moment Tensor (GCMT) and
Preliminary Determinations of Epicenters (Monthly Listing)
(PDE), see Section~\ref{catl}.
The number distributions obtained for these catalogs are
tested statistically to determine which of the theoretical
distribution fits them.
To do this we apply chi-square test to several subdivisions of
two catalogs, the test shows that the Poisson law can be
rejected as an approximation for most of the sub-catalogs
(Section~\ref{numb}).
To investigate goodness-of-fit of the NBD to various
catalog subdivisions we calculate upper statistical moments of
the number distributions (skewness and kurtosis) and compare
them to the potential values if these distributions follow the
Poisson or the NBD.
The comparison shows that for finer catalog time subdivision,
the NBD fails to fit empirical distributions
(Section~\ref{numb}).

Section~\ref{sim} studies to what extent the difference
between empirical and theoretical estimates of skewness and
kurtosis can be assigned to random fluctuations if the numbers
follow the NBD.
To achieve this we simulate NBD variables to see their random
fluctuations.
In the discussion in Section~\ref{disc} we present
recommendations for earthquake number testing, in particular,
we discuss in which sub-catalogs the Poisson or the NBD are
applicable, and propose to use empirical distributions for
many cases where these theoretical laws do not yield a good
approximation to the actual seismicity pattern.

\section {Theoretical distributions
}
\label{theo}

Two statistical distributions are used to approximate
the earthquake number structure.
The Poisson distribution is traditionally applied for this
purpose in engineering seismology and in present CSEP tests.
It has been long recognized that the Poisson law is a poor
approximation of seismicity occurrence.
One conventional way to treat this problem is to decluster an
earthquake catalogue as suggested by CSEP (Schorlemmer {\it et
al.}\ 2007).
But there are several declustering procedures, mostly based on
{\sl ad-hoc} rules.
Hence declustered catalogues are not unique and usually are
not fully reproducible.

The Poisson distribution has the probability function
of observing $k$ events as
\be
f \, (k) \ = \ { {\lambda^k \exp (-\lambda) }
\over {k! } } \, .
\label{EN_Eq1}
\ee
For this distribution its mean and variance are equal to its
rate $\lambda$.
The main problem in fitting the Poisson distribution to
earthquake data is over-dispersion of earthquake arrangement,
i.e., the variance of the earthquake process is usually much
higher than its rate (mean).

The discrete statistical distribution which allows for
over-dispersion is the NBD.
The most frequently used (we call it {\sl standard}) form of
the probability density function for the NBD
generalizes the Pascal distribution (Feller 1968, Eq.~VI.8.1;
Hilbe 2007, Eq.~5.19; Kozubowski \& Podg\'orski, 2009):
\ba
f \, (k) \ &=& \
{ { \tau \, (\tau + 1) \dots (\tau + k - 2) \, (\tau + k - 1)}
\over {k!} } \, \times \, \theta^\tau (1 - \theta)^k \ = \
{\tau + k - 1 \choose \tau -1} \, \times \, \theta^\tau (1 -
\theta)^k \
\nonumber\\
\ &=& \
{\tau + k - 1 \choose k} \, \times \, \theta^\tau (1 -
\theta)^k \ \ = \ \
{ {\Gamma (\tau + k) } \over {\Gamma (\tau) \, k! } } \,
\times \, \theta^\tau (1 - \theta)^k \, ,
\label{EN_Eq2}
\ea
where $ k = 0, 1, 2, \dots$, $\Gamma$ is the gamma function,
$ 0 \le \theta \le 1$, and $\tau > 0$, for the Pascal
distribution $\tau$ is a positive integer.

For $\theta \to 1$ and $\tau (1 - \theta) \to \lambda$
expression (\ref{EN_Eq2}) tends to (\ref{EN_Eq1}) (Feller,
1968, p.~281); the negative binomial distribution becomes the
Poisson distribution; the latter distribution is a special
case of the former.

\section {Earthquake catalogues
}
\label{catl}

To study the earthquake number distribution in an empirical
setting, we investigated it in sub-catalogs of two global
data-sets: global CMT catalog (GCMT) and PDE worldwide
catalog.
These catalogs have been selected because they are reasonably
uniform in coverage, location, magnitude, and time errors
(Kagan 2003).
Regional and local sub-sets of these catalogs can be also
studied to see their number distribution; local catalogs are
also useful for this purpose.
However, boundary conditions may strongly influence the number
distribution and such biases are difficult to take into
account.

We studied earthquake distributions and clustering for the
global CMT catalogue of moment tensor inversions compiled by
the CMT group (Ekstr\"om {\it et al.}\ 2012).
The present catalogue contains more than 45,000 earthquake
entries for the period 1977/1/1 to 2015/12/31.
The earthquake size is characterized by a scalar seismic
moment $M$.
The moment magnitude can be calculated from the seismic moment
(measured in Nm) value as
\be
m_W \ = \ (2/3) \cdot \log_{10} M - 6
\, .
\label{EN_Eq3}
\ee
The magnitude threshold for the catalogue is $m5.8$ (Kagan
2003).

The PDE worldwide catalogue is published by the USGS (U.S.\
Geological Survey 2008).
In its final form, the catalogue available at the time this
article was written ended on January 1, 2015.
The catalogue measures earthquake size, using several magnitude
scales, and provides the body-wave ($m_b$) and surface-wave
($M_S$) magnitudes for most moderate and large events since
1965 and 1968, respectively.
The moment magnitude ($m_W$) estimate has been added recently.
As the magnitude threshold Kagan \& Jackson (2016) propose
accepting $m_t = 5.0$.

\section {Earthquake numbers distribution testing
}
\label{numb}

In Table~\ref{Table1} we study whether the Poisson law can be
statistically rejected for various catalog subdivisions.
Since the Poisson distribution (\ref{EN_Eq1}) corresponds to
the NBD (\ref{EN_Eq2}) with the restriction $\theta \to
1.0$, i.e., there is no clustering, the statistical test
involves comparison of two log-likelihood values, the Poisson
($\ell_0$) and NBD ($\ell$).
Since the number of the degrees of freedom in these
distributions differs by one, the value of $ 2 \times (\ell -
\ell_0)$ should be distributed as $\chi^2_1$, i.e. the
chi-square distribution with one degree of freedom (Wilks
1962)
\be
\chi^2_1 \, (x) \ = \ 2 \times [ \, 1 - \Phi \, (\sqrt x \, )
\, ] \, ,
\label{EN_Eq4}
\ee
where $\Phi$ is the Gaussian cumulative distribution.

In this table we find that for most cases of interest the
Poisson distribution can be rejected statistically
at a high significance level in favor of the NBD.
Comparing upper two lines for North-West Pacific seismicity
demonstrates the influence of large earthquakes, in particular
the Tohoku 2011, $m=9.1$ event and its aftershocks on the
significance level.
Only for one case, South-West Pacific, the Poisson rejection
level (around 90\%) is less than 95\% which is usually
considered a threshold value.

Thereafter we investigate whether these distributions fit
observed distributions of seismicity.
In Tables~\ref{Table2} and \ref{Table3} we list earthquake
number properties of various sub-catalogs extracted from the
GCMT and PDE global data-sets.
In the sub-catalogs we modify catalogs magnitude threshold and
temporal subdivision: such data-sets can be used in earthquake
rate forecasts and testing.
In columns 5-8 of these tables parameter values for the
Poisson and NBD are listed; these parameters are determined by
using the two first moments of the number pattern.

It would be important to find out how these sub-catalogs fit
both theoretical distributions.
One way to accomplish this would be by applying the standard
statistical goodness-of-fit techniques, such as Cramer
von-Mises test (D'Augustino \& Stephens, 1986; Stephens,
1986).
However, such testing requires obtaining the number
distribution for each case, which would be a time-consuming
operation.

Goodness-of-fit could also be tested by using the sub-catalogs
statistical moment structure, which can be obtained in a
relatively simple way.
For this purpose we study the upper statistical moments of
the earthquake numbers, skewness and kurtosis
(Bowman \& Shenton, 1986; Li \& Papadopoulos, 2002),
and compare them to the theoretical values for both
distributions (Tables~\ref{Table2} and \ref{Table3}).
We calculate the values for skewness ($\eta$) and kurtosis
($\psi$) for theoretical statistical distributions, using
the formulas developed by Evans et al. (2000).

It is clear from these tables (\ref{Table2} and \ref{Table3})
that the empirical values for skewness and kurtosis increase
for the smaller magnitude threshold and increase with even
greater intensity for small temporal subdivision of catalogs.
As is known, the Poisson distribution for large rate values
approaches the Gaussian law, therefore its skewness and
kurtosis both tend to zero for large earthquake rates: for the
Gaussian law these values are identically zero.
Positive values for skewness mean that the distribution is
non-symmetric toward higher values of an argument.
For kurtosis positive values signify that the distribution has
heavy/fat tails (called lepto-kurtosis).

A calculation of the NBD skewness and kurtosis levels based on
the values of the first two statistical moments of the
distribution, shows rapid increase of these upper moments
levels.
However, the observed values of skewness and kurtosis,
especially for the PDE catalog (Table~\ref{Table3}), rise even
faster, indicating that for small time intervals the
earthquake number distribution is even more heavy-tailed than
the NBD expects.
The earthquake numbers for the GCMT catalog
(Table~\ref{Table2}) generally follows the same pattern, but
because of the higher magnitude threshold (5.8 vs 5.0 for PDE)
the number framework is not as obvious.

Kagan and Jackson (2016, Figs.~8-9) explored the annual
earthquake numbers and their statistical distribution as
compared to the Poisson and NBD laws for world-wide PDE
catalog.
The diagrams showed that the Poisson distribution is
inappropriate for the number pattern approximation, whereas
the NBD appears to fit the numbers reasonably well (compare
Table~\ref{Table1}).

Fig.~\ref{fig01} shows the annual earthquake numbers for the
South-West Pacific with 95\% levels calculated for the Poisson
and NBD laws.
Large earthquake numbers usually correspond to the occurrence
of a big mainshocks accompanied by an extensive aftershock
sequence.
Although in four cases the NBD levels are exceeded by observed
numbers, this feature is comparable to about 2 cases one
should expect for 95\% confidence levels.
The similar numbers for the Poisson distribution are
grossly excessive.

Fig.~\ref{fig02} demonstrates a fit of the empirical annual
number distribution for both theoretical curves -- Poisson and
NBD.
The NBD curve once again fits better than the Poisson curve,
however the difference between the fits is not large.

Figs.~\ref{fig03}--\ref{fig05} demonstrate
the approximation of the earthquake numbers in the PDE
catalog, subdivided into 1000 intervals of 16.8 days.
Earthquake numbers exhibit very large fluctuations with peaks
of activity exceeding 600 (see Fig.~\ref{fig03}).
Neither of the theoretical distributions approximates the
observed structure well in Fig.~\ref{fig04}; in
Fig.~\ref{fig05} the distribution's heavy tail for the upper
seismicity levels is clearly shown, even though the NBD curve
is far away from this tail.

These results mean that that NBD is not a good fit for
the earthquake number distribution at short time intervals.
Unfortunately, as statistical literature (Johnson et al.,
2005) demonstrates, there is none of the standard discrete
distributions having two degrees of freedom similarly to the
NBD.
There are many more complicated distributions, such as the
generalized Poisson or the generalized NBD, etc., but it is
unlikely that their application would be easy and yield
reproducible results.
Therefore to test earthquake numbers at small time intervals
we would need to apply new techniques.
One way to test earthquake number distribution
would be to use observed distributions for
each catalog.
These distributions, such as shown in Figs.~\ref{fig02},
\ref{fig04}, and \ref{fig05} would most likely need to be
smoothed during the testing.

\section {NBD Simulation
}
\label{sim}

To check our results displayed in Tables~\ref{Table2} and
\ref{Table3} we simulate the NBD and process the simulated
catalogs similar to real catalogs.
To simulate the NBD we use the procedure proposed by Evans et
al. (2000, see also Kozubowski \& Podg\'orski, 2009): we first
simulate a series of variables distributed according to the
geometric distribution (${\bf G :} \ \theta$).
The NBD simulation is obtained as a sum of $\tau$ geometric
variables
\be
{\bf NB :} \ \tau, \theta \ \sim \ \sum_{i=1}^\tau \,
({\bf G :} \ \theta)
\, ,
\label{EN_Eq5}
\ee
i.e., $\tau$ is here the integer, thus is effect we simulate
the Pascal distribution, that is a special case of the NBD.

The geometric distribution is simulated by a formula
\be
{\bf G :} \ \sim [ \, \log {\bf R}/ \log (1-\theta) \, ] - 1
\, .
\label{EN_Eq6}
\ee
where {\bf R} is the uniform random variate $1.0 \ge {\bf R} >
0.0$.
The result is rounded to the next larger integer (Evans et al.
2000, p.~108).

As simulation input we accept parameter values that are similar
to Table~\ref{Table3}, line 14: $\theta = 0.063$, $\tau = 4$.
The simulation results reproduce the value of the input
parameters, for example, $\hat \theta = 0.064 \pm 0.003$ and
$\hat \tau = 4.06 \pm 0.21$.
Similarly we obtain $\eta_n = 0.99 \pm 0.03$ and $\psi_n =
1.48 \pm 0.08$ as well as $\eta_p = 0.129 \pm 0.001$ and
$\psi_p = 0.0167 \pm 0.0003$.

Fig.~\ref{fig06} displays the scatterplot of skewness
($\eta$) and kurtosis ($\psi$) for the NBD parameters.
The values of $\eta$ and $\psi$ correspond well to those of
Table~\ref{Table3} (line 14); their standard deviations show
that we might expect large random fluctuations of these
variables, especially kurtosis: $\eta_s = 1.012 \pm 0.126$ and
$\psi_s = 1.48 \pm 0.627$.
The values of $\eta_s$ and $\psi_s$ and especially their
standard deviations are different from those shown in previous
paragraph for the NBD because the upper values are obtained
when the NBD parameters $\theta$ and $\tau$ are first
estimated from the simulation series, and their
skewness and kurtosis is calculated from these values (as was
done in Tables~\ref{Table2} and \ref{Table3}, columns 11 and
12).
However, $\eta_s$ and $\psi_s$ are evaluated directly from
the simulated series, thus they are equivalent to $\eta_o$ and
$\psi_o$ in Tables~\ref{Table2} and \ref{Table3} (columns 9
and 10).
However, the input catalog in the simulation case is produced
for the NBD whereas the original earthquake catalogs in the
Tables are processed.
By comparing both results we could see whether the earthquake
number distribution is similar in any way to the NBD.
Thus the significant difference between $\eta_s$ and $\eta_o$
as well as between $\psi_s$ and $\psi_o$ again demonstrates
that earthquake number distribution for this catalog temporal
subdivision is far from be approximated by the NBD.
However, the coefficient of variation for $\eta_s$ and
$\psi_s$ variables shown in Fig.~\ref{fig06} is 12\% and 42\%
respectively, $\eta_o$ and $\psi_o$ in Tables~\ref{Table2} and
\ref{Table3} are often vary to a much greater degree; for 1000
intervals $\eta_o = 7.628$ and $\psi_o = 110.5$ in
Table~\ref{Table3}.
As is seen from Fig.~\ref{fig06} the estimates $\eta_s$ and
$\psi_s$ are strongly correlated, with the correlation
coefficient $\rho = 0.919$.

To see how simulation results change if longer time intervals
are used to subdivide catalogs, we simulate the number
distribution for annual intervals.
Input parameters are $\theta = 0.015$, $\tau = 19$ (see
Table~\ref{Table3}).
The results as follows:
$\lambda = 1252.9 \pm 46.7$,
$\theta = 0.0159 \pm 0.0036$, $\tau = 20.18 \pm 4.50$;
$\eta_s = 0.45 \pm 0.43$, $\psi_s = 0.24 \pm 1.32$, and
$\rho = 0.804$;
$\eta_n = 0.453 \pm 0.047$, $\psi_n = 0.311 \pm 0.064$; and
$\eta_p = 0.0283 \pm 0.0005$, $\psi_p = 0.00080 \pm
0.00003$.
Whereas almost all the values correspond closely to those of
line 9 Table~\ref{Table3}, $\eta_o$ and $\psi_o$ are within
the random fluctuations of the simulated $\eta_s$ and
$\psi_s$.
Fig.~\ref{fig07} illustrates this point.

As a more detailed and accurate statistical test, we compare
the difference of two $\eta$ or $\psi$ values with their
standard deviations, $\sigma_\eta$ and $\sigma_\psi$.
The ratio
\be
z_\eta = {{\eta_s - \eta_o} \over {\sigma_\eta }}
\label{EN_Eq7}
\ee
is distributed for a large number of events ($n > 30 $)
according to a Gaussian distribution with a standard
deviation of 1.0.
In principle we need to use standard deviations for both
items compared, but we have only one deviation for simulated
catalogs, thus the test is only approximate.
We obtain $z_\eta = -2.07$ and $z_\psi = -0.84$.
Thus for skewness the hypothesis of both items equality is
rejected at the significance level slightly higher than
97.5\%, whereas for kurtosis the equality of both values is
not statistically rejected.
Fig.~\ref{fig07} can serve as a confirmation of the above
results: only two points of 100 are larger than $\eta_o$,
whereas eleven simulation points exceed $\psi_o$ value.

\section{Discussion }
\label{disc}

We reviewed the theoretical and statistical tools useful in
constructing earthquake number test in forecasts of seismic
activity as presently practiced by the CSEP.
These tools can be used by forecasts practitioners to produce
a practical testing algorithm for each particular earthquake
catalog and its subdivision.

Generally speaking, the Poisson distribution works reasonably
well for yearly samples of magnitude $> 6.5$.
The NBD works better for smaller event, and especially for
shorter time intervals, but neither of the theoretical
distributions is adequate for really small events and even
shorter intervals.

To see which statistical distribution is the most appropriate
for constructing the number test, we study statistical moments
of earthquake catalogs and sub-catalogs.
Statistical moments are relatively easy to investigate and
they are useful in characterizing earthquake occurrence
arrangement and seeing which theoretical or empirical laws are
most appropriate in approximating them.

We showed that three distributions are useful in number
testing: the Poisson, NBD, and the empirical distribution.
The Poisson distribution that was traditionally applied for
number testing can be only used in restricted cases of the
high magnitude threshold.
The NBD could be used in most cases for extended time
intervals forecasts.
However, if forecasts are considered for shorter time
intervals such as one month, weeks or days, the fluctuation of
earthquake numbers is such that no theoretical distribution
can reasonably fit their pattern.
Consequently, only empirical distributions are to be used for
this purpose.

\section*{7 ~ Conclusions }
\label{conc}

Our results on the forecast testing of earthquake number
distribution can be summarized as follows:
\hfil\break
$\bullet$ \quad
1) The Poisson distribution can be used for catalogs of large
earthquakes, with magnitude 6.5 and higher.
These earthquakes largely occur in a statistically
independent manner and their clustering, though present to a
minor degree in long catalogs, would not significantly change
the result.
\hfil\break
$\bullet$ \quad
2) The NBD can be used for testing distributions in
large earthquake catalogs and for long testing periods.
\hfil\break
$\bullet$ \quad
3) No standard statistical distribution appears to fit the
event number distribution for short time intervals, of the
order of weeks and days.
Thus we need to employ empirical distributions obtained for
each particular catalog.
To be useful such a distribution needs to be properly
smoothed.


%

%
\begin{table}
\caption{Testing earthquake number distribution
}
\vspace{10pt}
\label{Table1}
\begin{tabular}{lccrccrccc}
\hline
& & & & & & & & & \\[-25pt]
\multicolumn{1}{c}{Catalog/Region }&
\multicolumn{1}{c}{n }&
\multicolumn{1}{c}{$<N>$ }&
\multicolumn{1}{c}{$\sigma$(N) }&
\multicolumn{1}{c}{$\theta$ }&
\multicolumn{1}{c}{$\tau$ }&
\multicolumn{1}{c}{$\ell$ }&
\multicolumn{1}{c}{$\ell_0$ }&
\multicolumn{1}{c}{$2 \times \Delta \ell$ }&
\multicolumn{1}{c}{$\chi^2$ }
\\[5pt]
\hline
& & & & & & & & & \\[-15pt]
CMT NW 77-15 & 39 & 36.62 & 151.47 & 0.242 & 11.67 & $-9.66$ & $-31.64$ & 43.95 & 3.37e-11 \\
CMT NW 77-10 & 34 & 35.53 & 80.779 & 0.440 & 27.90 & $-6.74$ & $-11.03$ & 8.594 & 0.0034 \\
CMT SW 77-15 & 39 & 60.54 & 106.15 & 0.570 & 80.36 & $-5.51$ & $-6.859$ & 2.694 & 0.1007 \\
CMT GL 77-15 & 39 & 177.18 & 737.33 & 0.240 & 56.04 & $-7.07$ & $-15.25$ & 16.36 & 5.23e-05 \\
PDE GL 69-14 & 46 & 1280.6 & 88191 & 0.0145 & 18.87 & $-10.2$ & $-252.9$ & 485.5 & 0 \\
PDE GL 69-03 & 35 & 1147.0 & 16208 & 0.0708 & 87.35 & $-7.09$ & $-21.84$ & 29.50 & 5.59e-08 \\
& & & & & & & \\[-15pt]
\hline
\end{tabular}

\bigskip
CMT - Global Centroid-Moment-Tensor catalog:
NW - North-West Pacific;
SW - South-West Pacific;
GL - Global catalog.
PDE - Preliminary Determinations of Epicenters catalog.
n - number of annual intervals;
$<N>$ - average annual number of earthquakes;
$\sigma$(N) - standard deviation of N;
$\theta$ - clustering parameter of negative binomial
distribution (NBD);
$\tau$ - parameter of NBD;
$\ell$ - NBD log-likelihood;
$\ell_0$ - Poisson log-likelihood;
$\Delta(\ell) = \ell - \ell_0$ - log-likelihood difference;
$\chi^2$ - chi-square value.

\hfil\break
\vspace{5pt}
\end{table}

\newpage

\clearpage

%
%
%
\renewcommand{\baselinestretch}{1.25}
\setlength{\textheight}{22cm}
\tolerance=5000
\font\scaps=cmcsc10    
\font\bbf=cmbx10 scaled\magstep4
\font\rmbig=cmr10 scaled\magstep2
\font\bigtensl=cmsl10 scaled\magstep2

\oddsidemargin=-1.5cm
\evensidemargin=-1.5cm

%

\begin{table}
\caption{Values of NBD parameters and skewness and kurtosis for
various subdivisions of the 1977-2015 CMT catalogue, $m_{\rm
max} = 9.15$
}
\vspace{10pt}
\label{Table2}
\begin{tabular}{rcrrrcccrrccccr}
\hline
& & & & & & & & & & & & & & \\[-15pt]
\multicolumn{1}{c}{\#}&
\multicolumn{1}{c}{$m_t$}&
\multicolumn{1}{c}{$n$}&
\multicolumn{1}{c}{$N$}&
\multicolumn{1}{c}{$\lambda$}&
\multicolumn{1}{c}{$\sigma$}&
\multicolumn{1}{c}{$\theta$}&
\multicolumn{1}{c}{$\tau$}&
\multicolumn{1}{c}{$\eta_o$}&
\multicolumn{1}{c}{$\psi_o$}&
\multicolumn{1}{c}{$\eta_n$}&
\multicolumn{1}{c}{$\psi_n$}&
\multicolumn{1}{c}{$\eta_p$}&
\multicolumn{1}{c}{$\psi_p$}&
\multicolumn{1}{c}{$\Delta T$}
\\[2pt]
\hline
\multicolumn{1}{c}{1}&
\multicolumn{1}{c}{2}&
\multicolumn{1}{c}{3}&
\multicolumn{1}{c}{4}&
\multicolumn{1}{c}{5}&
\multicolumn{1}{c}{6}&
\multicolumn{1}{c}{7}&
\multicolumn{1}{c}{8}&
\multicolumn{1}{c}{9}&
\multicolumn{1}{c}{10}&
\multicolumn{1}{c}{11}&
\multicolumn{1}{c}{12}&
\multicolumn{1}{c}{13}&
\multicolumn{1}{c}{14}&
\multicolumn{1}{c}{15}
\\[2pt]
\hline
& & & & & & & & & & & & & & \\[-15pt]
1 & 7.0 & 416 & 39 & 10.67 & 10.22 & $-$    & $-$   & 0.219 & 0.070 & 0.287 & 0.074 & 0.306 & 0.094 & 365.2 \\
2 & 6.5 & 1340 & 39 & 34.36 & 54.33 & 0.632 & 59.1 & 0.086 & $-$0.254 & 0.293 & 0.120 & 0.171 & 0.029 & 365.2 \\
3 & 6.0 & 4359 & 39 & 111.8 & 348.0 & 0.321 & 52.9 & 0.556 & $-$0.171 & 0.280 & 0.116 & 0.095 & 0.009 & 365.2 \\
4 & 5.8 & 6910 & 39 & 177.2 & 742.5 & 0.239 & 55.5 & 0.712 & 0.017 & 0.271 & 0.109 & 0.075 & 0.006 & 365.2 \\
& & & & & & & & & \\
5 & 5.8 & 6910 & 5 & 1382 & 17096 & 0.081 & 121 & 0.352 & $-$1.319 & 0.182 & 0.049 & 0.027 & 0.001 & 2848.8 \\
6 & 5.8 & 6910 & 10 & 691.0 & 6857 & 0.101 & 77.4 & 0.762 & $-$0.647 & 0.228 & 0.078 & 0.038 & 0.002 & 1424.4 \\
7 & 5.8 & 6910 & 20 & 345.5 & 2106 & 0.164 & 67.8 & 0.990 & 0.468 & 0.244 & 0.089 & 0.054 & 0.003 & 712.2 \\
8 & 5.8 & 6910 & 39 & 177.2 & 742.5 & 0.239 & 55.5 & 0.712 & 0.017 & 0.271 & 0.109 & 0.075 & 0.006 & 365.2 \\
9 & 5.8 & 6910 & 50 & 138.2 & 567.0 & 0.244 & 44.5 & 0.543 & 0.041 & 0.303 & 0.137 & 0.085 & 0.007 & 284.9 \\
10 & 5.8 & 6910 & 100 & 69.10 & 229.3 & 0.301 & 29.8 & 0.609 & 0.220 & 0.372 & 0.206 & 0.120 & 0.015 & 142.4 \\
11 & 5.8 & 6910 & 200 & 34.55 & 91.33 & 0.378 & 21.0 & 1.055 & 2.376 & 0.449 & 0.296 & 0.170 & 0.029 & 71.2 \\
12 & 5.8 & 6910 & 500 & 13.82 & 30.26 & 0.457 & 11.6 & 1.566 & 7.305 & 0.614 & 0.550 & 0.269 & 0.072 & 28.5 \\
13 & 5.8 & 6910 & 1000 & 6.910 & 13.29 & 0.520 & 7.48 & 1.656 & 7.508 & 0.781 & 0.877 & 0.380 & 0.145 & 14.2 \\
14 & 5.8 & 6910 & 7122 & 0.970 & 1.34 & 0.723 & 2.53 & 1.899 & 7.234 & 1.524 & 3.113 & 1.015 & 1.031 & 2.0 \\
15 & 5.8 & 6910 & 14244 & 0.485 & 0.62 & 0.788 & 1.80 & 2.364 & 11.38 & 1.961 & 4.956 & 1.436 & 2.061 & 1.0 \\
& & & & & & & & & \\[-15pt]
\hline
\end{tabular}

\bigskip
$m_t$ is magnitude threshold value,
$n$ is the number of earthquakes,
$N$ is the number of time intervals,
$\lambda$ - earthquake rate of occurrence;
$\sigma$ - standard deviation of earthquake numbers;
$\theta$ - clustering parameter of negative binomial
distribution (NBD);
$\tau$ - parameter of NBD;
$\eta_o$, $\psi_o$ - observed skewness and kurtosis;
$\eta_n$, $\psi_n$ - skewness and kurtosis calculated for NBD;
$\eta_p$, $\psi_p$ - skewness and kurtosis calculated for
Poisson distribution;
$\Delta T$ -- interval duration in days;
$m_{\rm max}$ -- maximum observed magnitude.

\hfil\break
\vspace{5pt}
\end{table}

\newpage

\clearpage

%
%
\setlength{\textwidth}{19.5cm}
\renewcommand{\baselinestretch}{1.25}
\setlength{\textheight}{22cm}
\tolerance=5000
\font\scaps=cmcsc10    
\font\bbf=cmbx10 scaled\magstep4
\font\rmbig=cmr10 scaled\magstep2
\font\bigtensl=cmsl10 scaled\magstep2

\oddsidemargin=-1.75cm
\evensidemargin=-1.75cm

%



\begin{table}
\caption{Values of NBD parameters and skewness and kurtosis for
various subdivisions of the 1969-2014 PDE catalogue, $m_{\rm
max} = 9.0$
}
\vspace{10pt}
\label{Table3}
\begin{tabular}{rcrrrcccrrccccr}
\hline
& & & & & & & & & & & & & & \\[-15pt]
\multicolumn{1}{c}{\#}&
\multicolumn{1}{c}{$m_t$}&
\multicolumn{1}{c}{$n$}&
\multicolumn{1}{c}{$N$}&
\multicolumn{1}{c}{$\lambda$}&
\multicolumn{1}{c}{$\sigma$}&
\multicolumn{1}{c}{$\theta$}&
\multicolumn{1}{c}{$\tau$}&
\multicolumn{1}{c}{$\eta_o$}&
\multicolumn{1}{c}{$\psi_o$}&
\multicolumn{1}{c}{$\eta_n$}&
\multicolumn{1}{c}{$\psi_n$}&
\multicolumn{1}{c}{$\eta_p$}&
\multicolumn{1}{c}{$\psi_p$}&
\multicolumn{1}{c}{$\Delta T$}
\\[2pt]
\hline
\multicolumn{1}{c}{1}&
\multicolumn{1}{c}{2}&
\multicolumn{1}{c}{3}&
\multicolumn{1}{c}{4}&
\multicolumn{1}{c}{5}&
\multicolumn{1}{c}{6}&
\multicolumn{1}{c}{7}&
\multicolumn{1}{c}{8}&
\multicolumn{1}{c}{9}&
\multicolumn{1}{c}{10}&
\multicolumn{1}{c}{11}&
\multicolumn{1}{c}{12}&
\multicolumn{1}{c}{13}&
\multicolumn{1}{c}{14}&
\multicolumn{1}{c}{15}
\\[2pt]
\hline
& & & & & & & & & & & & & & \\[-15pt]
1 & 7.0 & 560 & 46 & 12.17 & 14.10 & 0.863 & 76.9 & 0.517 & 0.753 & 0.351 & 0.149 & 0.287 & 0.0821 & 365.2 \\
2 & 6.5 & 1635 & 46 & 35.54 & 58.12 & 0.612 & 56.0 & 0.212 & 0.286 & 0.298 & 0.124 & 0.168 & 0.0281 & 365.2 \\
3 & 6.0 & 4826 & 46 & 104.9 & 526.6 & 0.199 & 26.1 & 0.763 & 0.273 & 0.394 & 0.232 & 0.098 & 0.0095 & 365.2 \\
4 & 5.5 & 16651 & 46 & 362.0 & 6369 & 0.057 & 21.8 & 0.893 & 0.230 & 0.428 & 0.275 & 0.053 & 0.0028 & 365.2 \\
5 & 5.0 & 58909 & 46 & 1281 & 88191 & 0.015 & 18.9 & 1.386 & 1.436 & 0.460 & 0.318 & 0.028 & 0.0008 & 365.2 \\
& & & & & & & & & \\
6 & 5.0 & 58909 & 5 & 11781 & 387e4 & 0.003 & 35.9 & 1.452 & 0.187 & 0.334 & 0.167 & 0.009 & 0.0001 & 3360 \\
7 & 5.0 & 58909 & 10 & 5891 & 121e4 & 0.005 & 28.6 & 1.030 & 0.030 & 0.374 & 0.210 & 0.013 & 0.0002 & 1680 \\
8 & 5.0 & 58909 & 20 & 2945 & 400e3 & 0.007 & 21.8 & 1.088 & $-$0.069 & 0.428 & 0.275 & 0.018 & 0.0003 & 840.0 \\
9 & 5.0 & 58909 & 46 & 1281 & 88191 & 0.015 & 18.9 & 1.386 & 1.436 & 0.460 & 0.318 & 0.028 & 0.0008 & 365.2 \\
10 & 5.0 & 58909 & 50 & 1178 & 88856 & 0.013 & 15.8 & 1.712 & 2.902 & 0.503 & 0.379 & 0.029 & 0.0009 & 336.0 \\
11 & 5.0 & 58909 & 100 & 589.1 & 27994 & 0.021 & 12.7 & 2.375 & 8.006 & 0.562 & 0.474 & 0.041 & 0.0017 & 168.0 \\
12 & 5.0 & 58909 & 200 & 294.5 & 8559 & 0.034 & 10.5 & 2.697 & 12.34 & 0.617 & 0.572 & 0.058 & 0.0034 & 84.0 \\
13 & 5.0 & 58909 & 500 & 117.8 & 2339 & 0.050 & 6.25 & 4.756 & 41.18 & 0.800 & 0.960 & 0.092 & 0.0085 & 33.6 \\
14 & 5.0 & 58909 & 1000 & 58.91 & 942.4 & 0.063 & 3.92 & 7.628 & 110.5 & 1.010 & 1.529 & 0.130 & 0.0170 & 16.8 \\
15 & 5.0 & 58909 & 8400 & 7.013 & 48.92 & 0.143 & 1.17 & 19.64 & 823.0 & 1.852 & 5.133 & 0.378 & 0.1426 & 2.0 \\
16 & 5.0 & 58909 & 16801 & 3.506 & 17.94 & 0.196 & 0.85 & 23.18 & 1224 & 2.180 & 7.100 & 0.534 & 0.2852 & 1.0 \\
& & & & & & & & & \\[-15pt]
\hline
\end{tabular}

\bigskip
$m_t$ is magnitude threshold value,
$n$ is the number of earthquakes,
$N$ is the number of time intervals,
$\lambda$ - earthquake rate of occurrence;
$\sigma$ - standard deviation of earthquake numbers;
$\theta$ - clustering parameter of negative binomial
distribution (NBD);
$\tau$ - parameter of NBD;
$\eta_o$, $\psi_o$ - observed skewness and kurtosis;
$\eta_n$, $\psi_n$ - skewness and kurtosis calculated for NBD;
$\eta_p$, $\psi_p$ - skewness and kurtosis calculated for
Poisson distribution;
$\Delta T$ -- interval duration in days;
$m_{\rm max}$ -- maximum observed magnitude.

\hfil\break
\vspace{5pt}
\end{table}

\newpage

\clearpage

\setlength{\oddsidemargin}{0cm}
\setlength{\evensidemargin}{0cm}

\subsection* {Acknowledgments
}
\label{Ackn}

I am grateful to Peter Bird of UCLA for useful discussions.
The author appreciates partial support from the National
Science Foundation through grant EAR-1045876, as well as from
the Southern California Earthquake Center (SCEC).
This research was supported by the Southern California 
Earthquake Center. SCEC is funded by 
NSF Cooperative Agreement EAR-1033462 \& USGS Cooperative 
Agreement G12AC20038. 
Publication 7071, SCEC.

\pagebreak

\renewcommand{\baselinestretch}{1.75}
\setlength{\textheight}{22cm}
\oddsidemargin=0.0cm
\evensidemargin=0.0cm

\def\reference{\hangindent=22pt\hangafter=1}

\centerline { {\sc References} }
\vskip 0.1in
\parskip 1pt
\parindent=1mm
\def\reference{\hangindent=22pt\hangafter=1}

\reference
Bird, P., D. D. Jackson, Y. Y. Kagan, C. Kreemer, and R. S.
Stein, 2015.
GEAR1: a Global Earthquake Activity Rate model constructed
from geodetic strain rates and smoothed seismicity,
{\sl Bull.\ Seismol.\ Soc.\ Amer.}, {\bf 105}(5),
2538-2554, doi:10.1785/0120150058, (plus electronic
supplement).

\reference
Bowman, K. O., Shenton, L. R. (1986).
Moment (${\sqrt b_1}, b_2$) Techniques.
In: D'Augustino, R. B., Stephens, M. A., eds. Goodness of Fit
Techniques. Statistics: Textbooks and Monographs. Vol. 68. New
York: Marcel Dekker, pp. 279-327.

\reference
D'Augustino, R. B., and M. A. Stephens (1986).
Goodness-of-fit techniques, in Statistics: Textbooks and
Monographs, Vol. 68, Marcel Dekker, New York, 560 pp.

\reference
Ekstr\"om, G., M. Nettles \& A.M. Dziewonski, 2012.
The global CMT project 2004-2010: Centroid-moment tensors for
13,017 earthquakes,
{\sl Phys.\ Earth Planet.\ Inter.}, {\bf 200-201}, 1-9.
10.1016/j.pepi.2012.04.002

\reference
Evans, M., N. Hastings, and B. Peacock, 2000.
{\sl Statistical Distributions},
3rd ed., New York, J. Wiley, 221~pp.

\reference
Feller, W., 1968.
{\sl An Introduction to Probability Theory and its
Applications}, {\bf 1}, 3-rd ed., J. Wiley, New York, 509~pp.

\reference
Johnson, N.\ L., A. W. Kemp, and S. Kotz, 2005.
{\sl Univariate Discrete Distributions},
3rd ed., Wiley, Hoboken, New Jersey, 646~pp.

\reference
Kagan, Y. Y., 2003.
Accuracy of modern global earthquake catalogs,
{\sl Phys.\ Earth Planet.\ Inter.\ (PEPI)}, {\bf 135}(2-3),
173-209.

\reference
Kagan, Y. Y., 2010.
Statistical distributions of earthquake numbers:
consequence of branching process,
{\sl Geophys.\ J. Int.}, {\bf 180}(3), 1313-1328.
doi: 10.1111/j.1365-246X.2009.04487.x

\reference
Kagan, Y. Y., 2014.
{\bf EARTHQUAKES: Models, Statistics, Testable Forecasts},
Hoboken, NJ, John Wiley \& Sons, 306~pp,
ISBN: 978-1118637913.

\reference
Kozubowski, T. J., and K. Podg\'orski, 2009.
Distributional properties of the negative binomial L\'evy
process,
{\sl Probability and Mathematical Statistics}, {\bf 29}(1),
43-71.

\reference
Li, G., and A. Papadopoulos, 2002.
A note on goodness of fit test using moments,
{\sl Statistica}, {\bf 62}(1), 71-86.

\reference
Stephens, M. A., 1986.
Tests based on EDF statistics,
in: Goodness-of-Fit Techniques, Statistics: Textbooks and
Monographs, Vol. 68, edited by: D'Augustino, R. B. and
Stephens, M. A., Marcel Dekker, New York, 97-194, 1986.

\reference
U.S.\ Geological Survey, {\sl Preliminary Determinations of
Epicenters (PDE)}, 2008.
U.S.\ Dep.\ of Inter., Natl.\ Earthquake Inf.\ Cent.
http://neic.usgs.gov/neis/epic/code\_catalog.html

\reference
Wilks, S.\ S., 1962.
{\sl Mathematical Statistics},
John Wiley, New York, 644~pp.

\clearpage

\newpage

\begin{figure}
\begin{center}
\includegraphics[width=0.60\textwidth,angle=0]{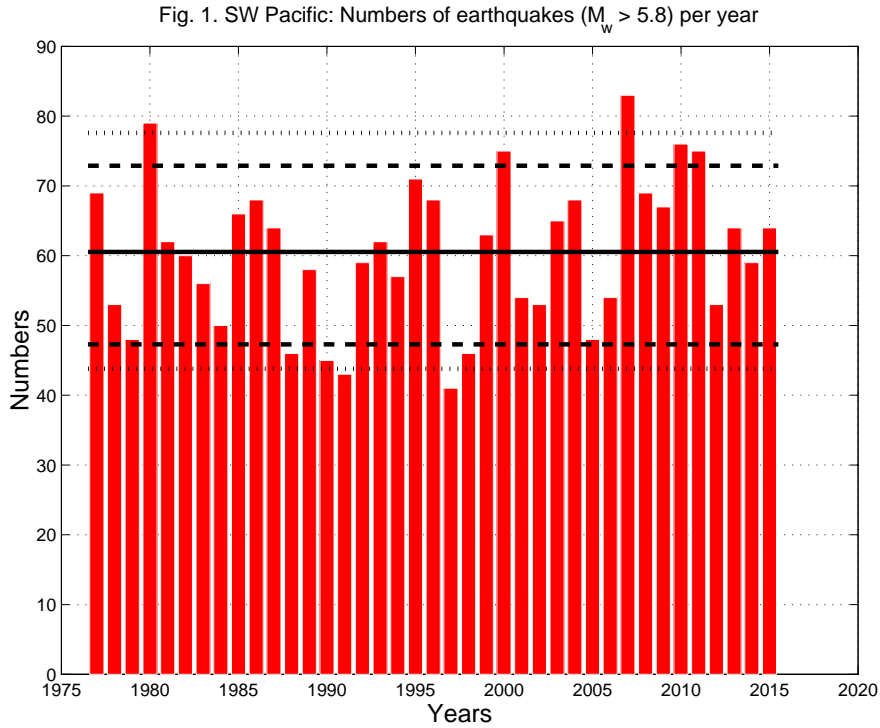}
\caption{\label{fig01}
Annual numbers of earthquakes for the GCMT catalog, 1977-2015,
$m \ge 5.8$, South-West Pacific.
Blue horizontal line shows average annual earthquake number,
two blue lines demonstrate 95\% confidence areas for the
Poisson (dashed lines) and NBD (dotted lines) distributions.
Only four annual earthquake numbers are outside of the NBD
confidence intervals.
For 95\% confidence one should expect that about 5\% of 39
entries would exceed the limits.
}
\end{center}
\end{figure}

\begin{figure}
\begin{center}
\includegraphics[width=0.60\textwidth,angle=0]{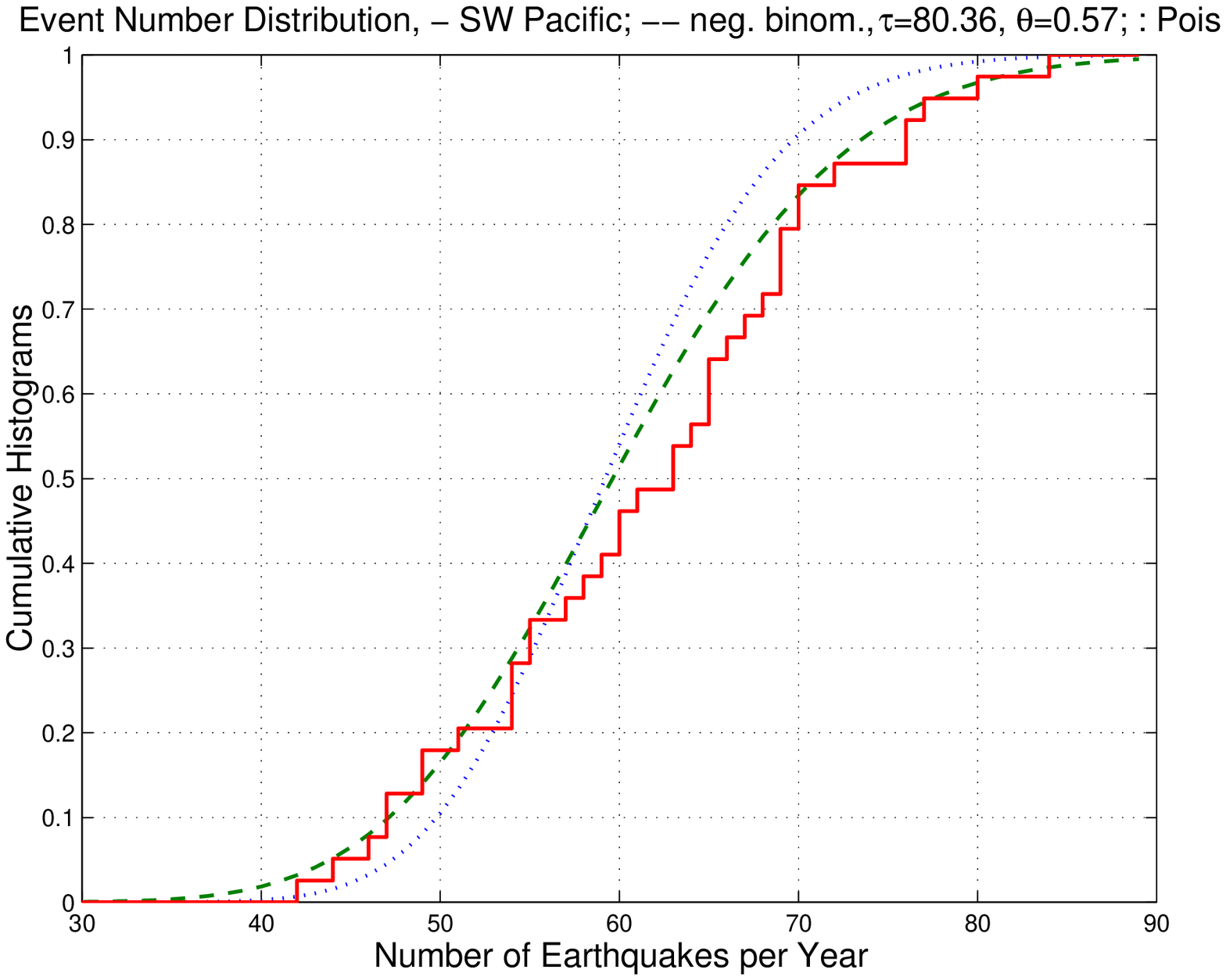}
\caption{\label{fig02}
}
\end{center}
Cumulative distribution of yearly earthquake numbers
for the GCMT catalog, 1977-2015, $m \ge 5.8$, South-West
Pacific.
The step-function shows the observed distribution, the dashed
curve is the theoretical Poisson distribution for $\lambda =
60.54$ and
the dashed curve is the fitted negative-binomial curve for
$\theta = 0.570$ and $\tau = 80.36$.
As expected from Table~\ref{Table1} results
the negative-binomial curve has a better fit than the Poisson
curve, though the difference is not large.
\end{figure}

\begin{figure}
\begin{center}
\includegraphics[width=0.60\textwidth,angle=0]{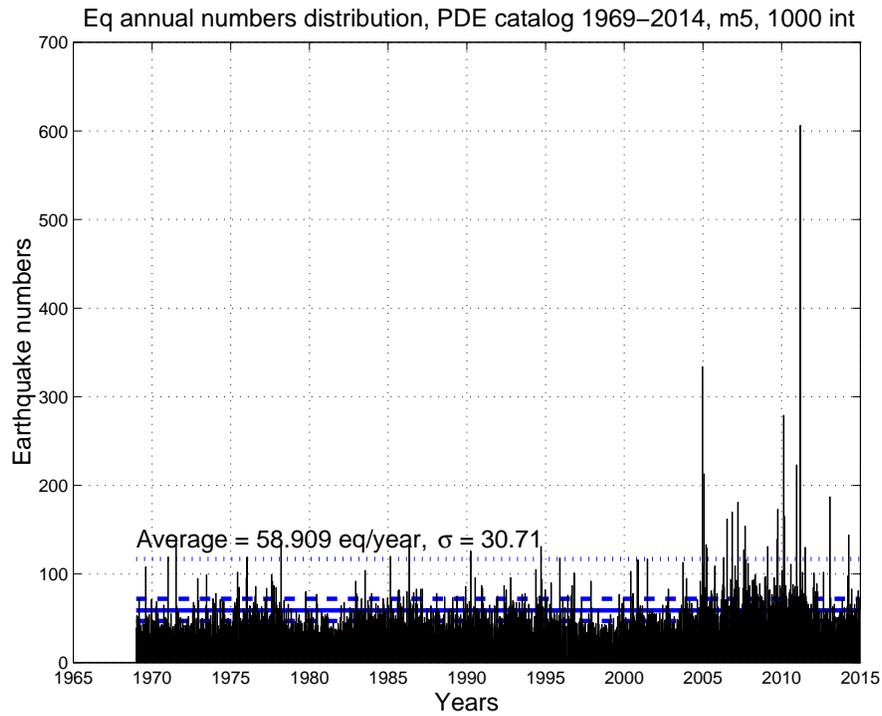}
\caption{\label{fig03}
Numbers of earthquakes for the PDE catalog, 1969-2014,
$m \ge 5.0$, subdivided into 1000 intervals (duration 16.8
days, see Table~\ref{Table3}).
Blue horizontal line shows average annual earthquake number,
two blue lines demonstrate 95\% confidence areas for the
Poisson (dashed lines) and NBD (dotted lines) distributions.
Many of the annual earthquake numbers are outside of the NBD
confidence intervals, thus the NBD poorly approximating the
number distribution.
}
\end{center}
\end{figure}

\begin{figure}
\begin{center}
\includegraphics[width=0.60\textwidth,angle=0]{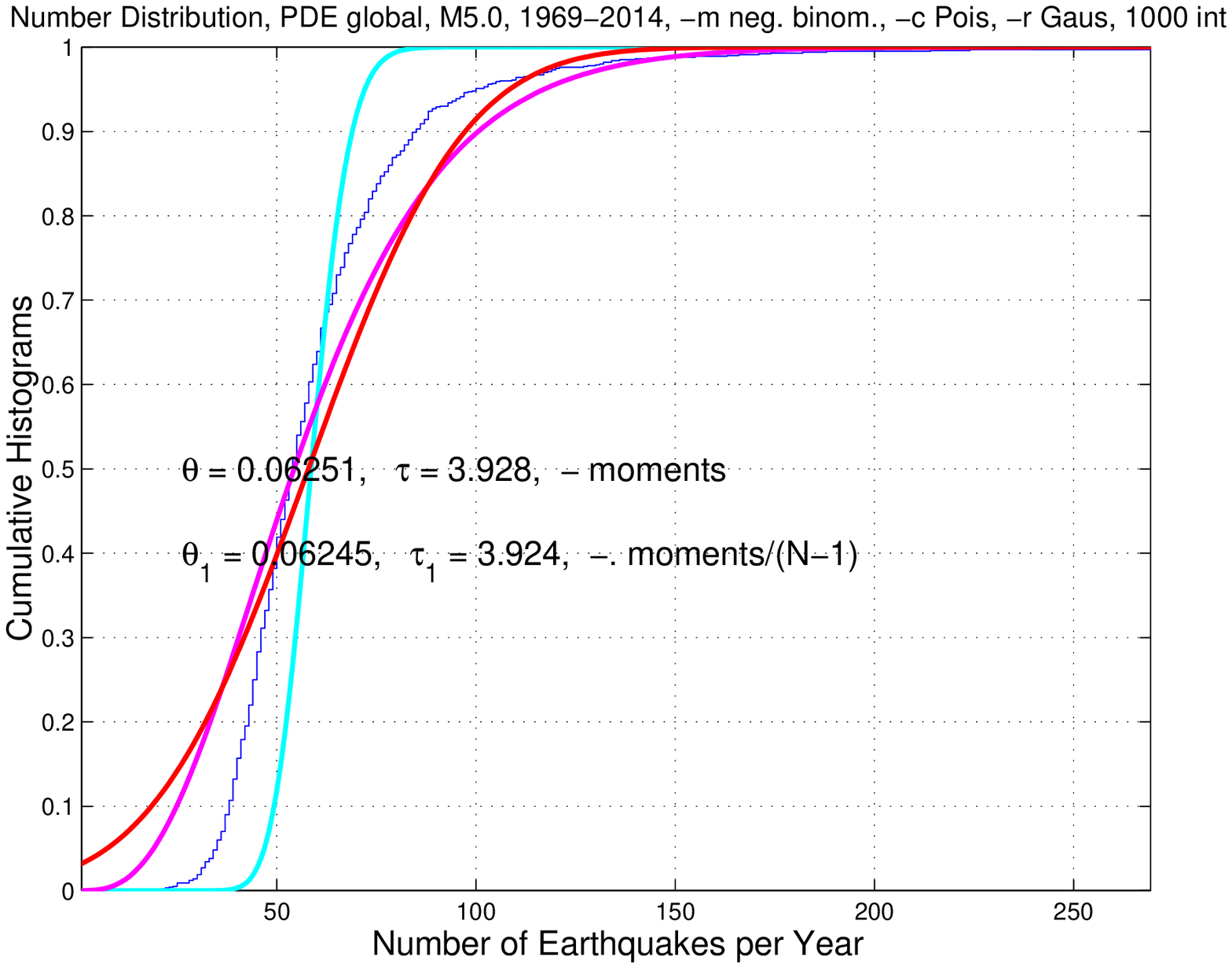}
\caption{\label{fig04}
}
\end{center}
Blue curve is cumulative distribution of yearly global
earthquake numbers for the PDE catalog, 1969-2014, $m \ge
5.0$.
The step-function shows the observed distribution, the red
curve is the Gaussian distribution, the cyan curve is the
theoretical Poisson distribution for $\lambda = 58.9$ and the
magenta solid curve is the fitted negative-binomial curve for
$\theta = 0.063$ and $\tau = 3.93$.
The negative-binomial curve has a better fit than the Poisson
curve.
\end{figure}

\begin{figure}
\begin{center}
\includegraphics[width=0.60\textwidth,angle=0]{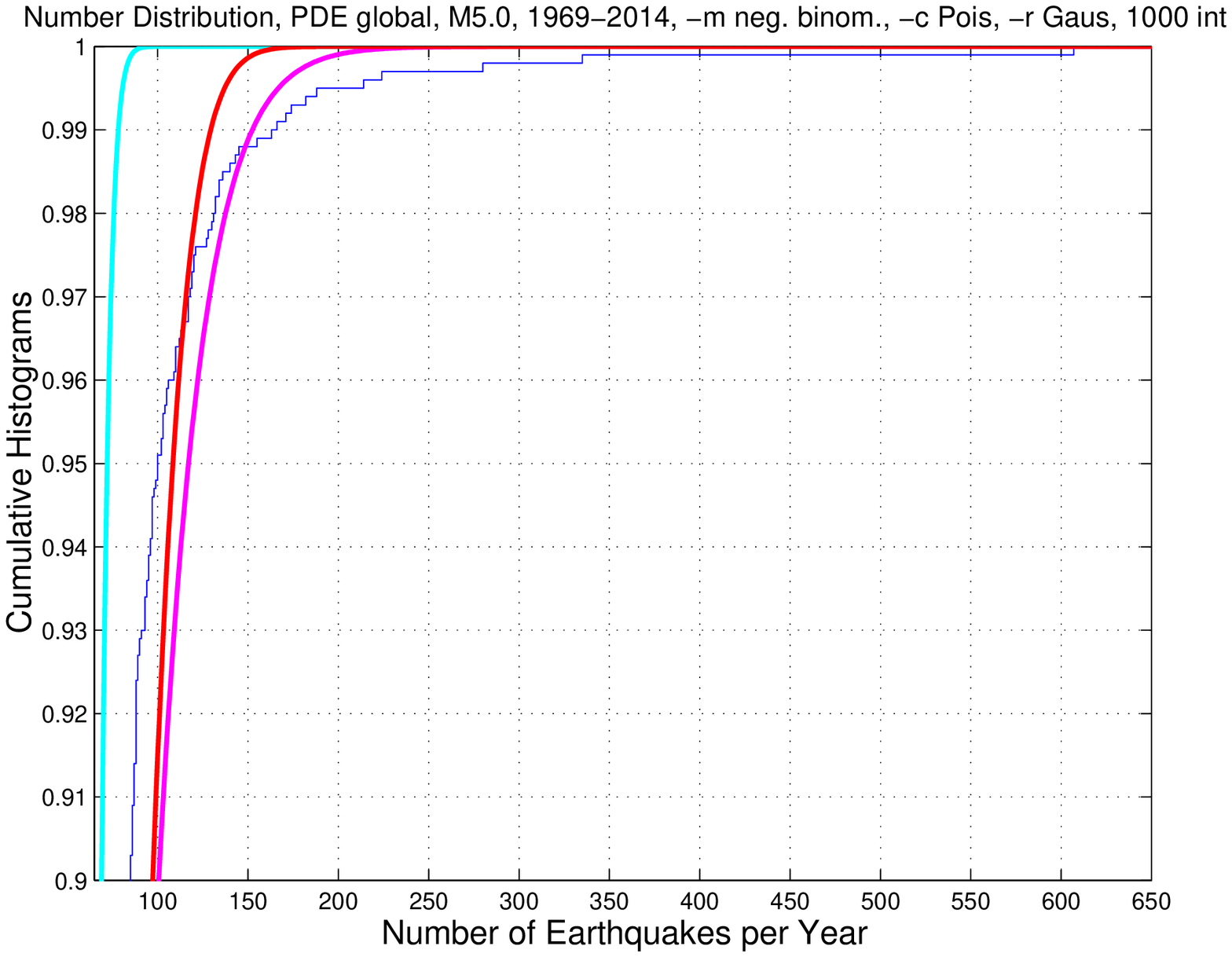}
\caption{\label{fig05}
}
\end{center}
Same as Fig.~\ref{fig04} but only the upper part of
distribution curves is displayed.
Kurtosis values for curves, from left to right (see
Table~\ref{Table3}):
$\psi_p = 0.02$,
$\psi_g = 0$,
$\psi_n = 1.53$,
$\psi_o = 110.5$.
\end{figure}

\begin{figure}
\begin{center}
\includegraphics[width=0.60\textwidth,angle=0]{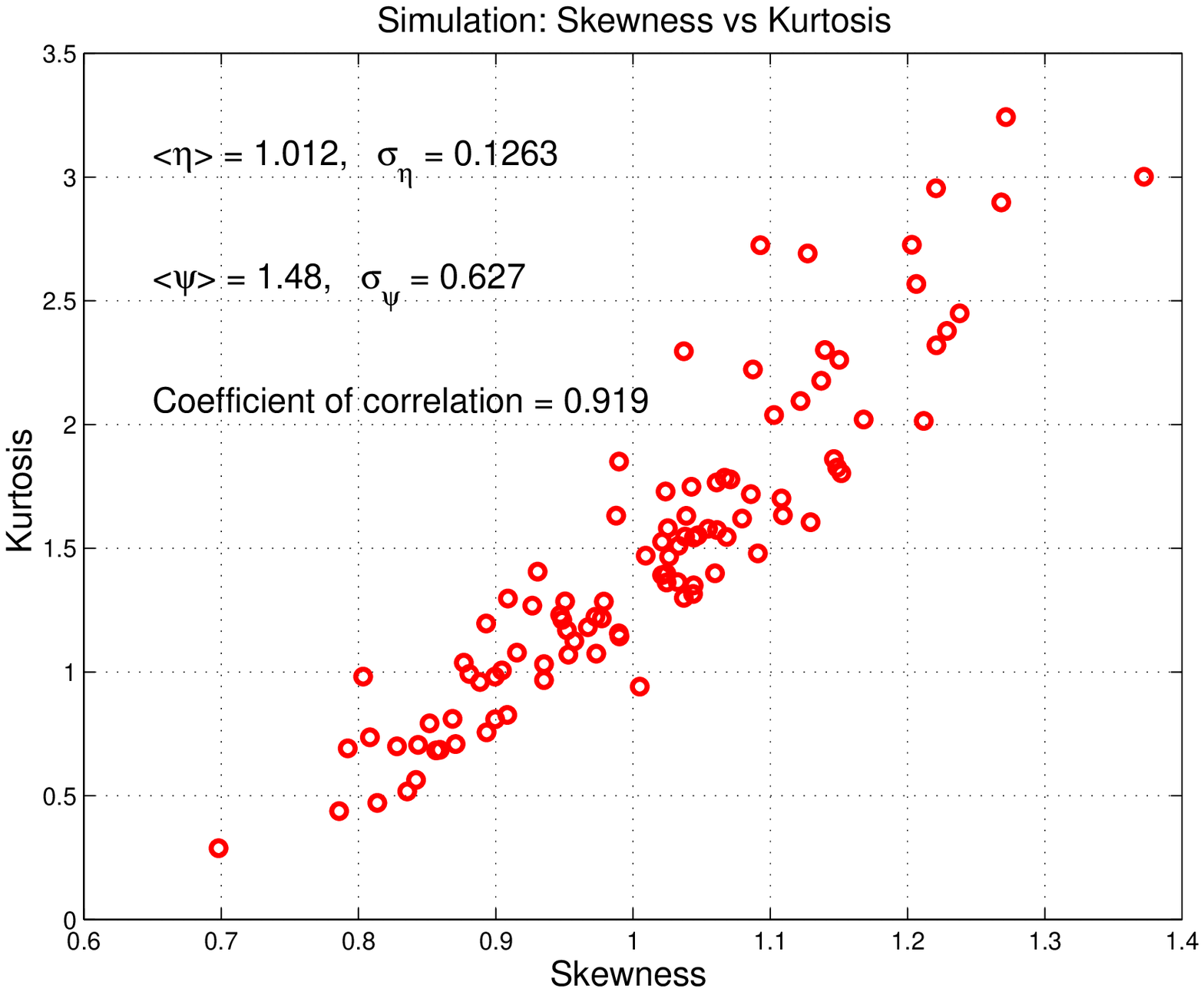}
\caption{\label{fig06}
}
\end{center}
Scatterplot of skewness ($\eta$) and kurtosis ($\psi$) for
simulated NBD for 1000 intervals catalog subdivision.
\end{figure}

\begin{figure}
\begin{center}
\includegraphics[width=0.60\textwidth,angle=0]{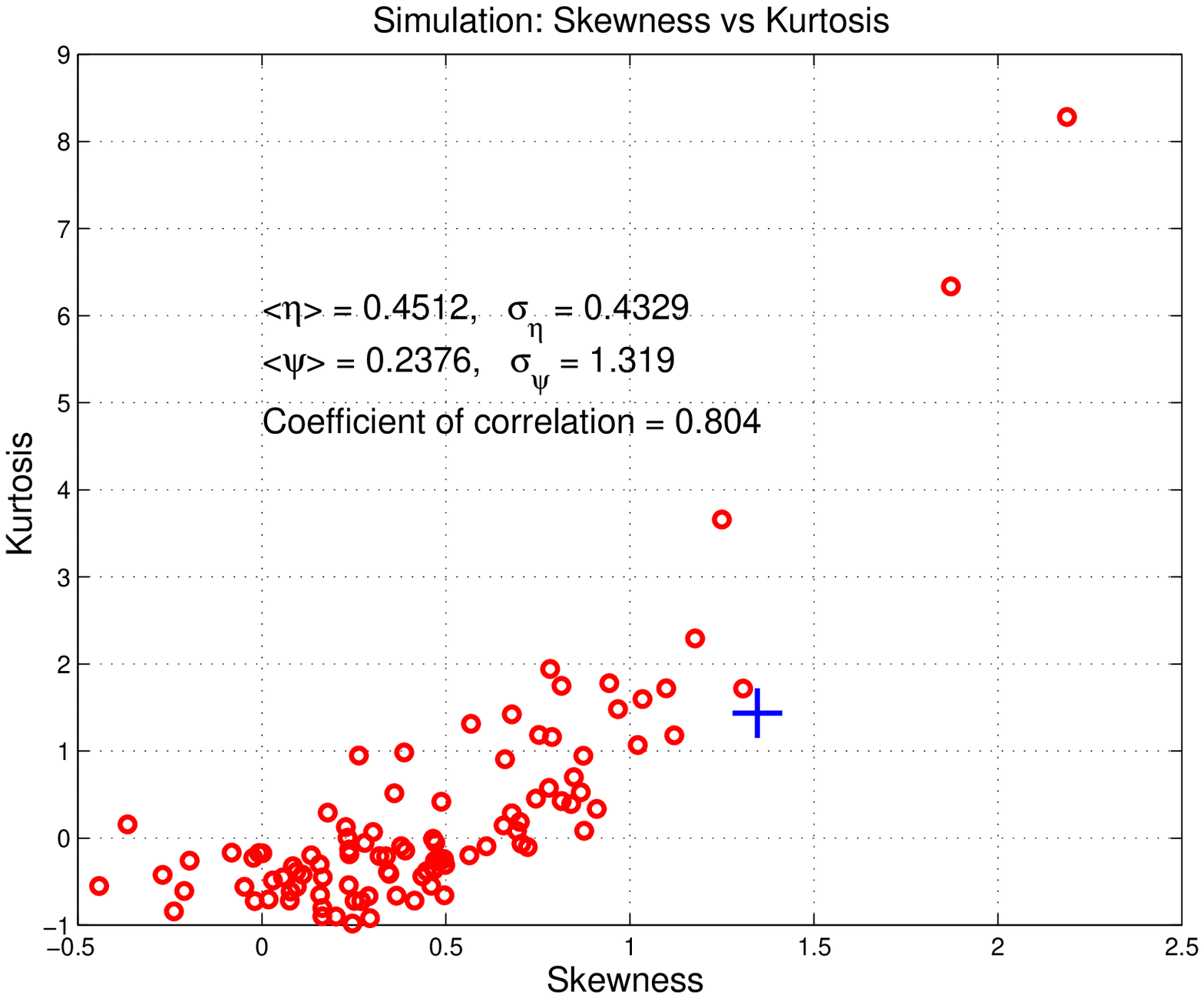}
\caption{\label{fig07}
}
\end{center}
Scatterplot of skewness ($\eta$) and kurtosis ($\psi$) for
simulated NBD for annual interval catalog subdivision.
Large blue cross shows skewness ($\eta_o$) and kurtosis
($\psi_o$) for the annual subdivision of the PDE catalog (see
Table~\ref{Table3}).
\end{figure}

\end{document}